\documentstyle[12pt]{article}
\begin{document}
\title{Relativistic nuclear recoil corrections to the  energy levels
 of hydrogen-like and high $Z$ lithium-like atoms
  in all orders in $\alpha Z$}
\author{A.N.Artemyev, V.M.Shabaev, and V.A.Yerokhin\\
{\it Department of Physics, St.Petersburg State University,}\\
{\it Oulianovskaya 1, Petrodvorets, St.Petersburg 198904, Russia}}
\maketitle
\begin{abstract}
The relativistic nuclear recoil corrections to
 the energy levels of low-laying states
 of hydrogen-like and high $Z$ lithium-like atoms in all orders in
$\alpha Z$  are calculated.
The calculations are carried out using the B-spline method for the
Dirac equation.
  For low $Z$ the results
of the calculation are in good agreement with the
$\alpha Z$ -expansion results.
 It is found that the nuclear recoil contribution,
 additional to the Salpeter's one,
to the Lamb shift ($n=2$) of hydrogen is $-1.32(6)\,kHz$.
The  total nuclear
recoil correction to the energy of the $(1s)^{2}2p_{\frac{1}{2}}-
(1s)^{2}2s$ transition in lithium-like
uranium constitutes $-0.07\,eV$ and is largely made up of
 QED contributions.

PACS number(s): 12.20.Ds, 31.30.Jv
\end{abstract}
\newpage
\section{Introduction}

As is known, in the non-relativistic approximation the nuclear recoil
correction for a hydrogen-like atom can be taken into account by using
the reduced mass $\mu=\frac{mM}{m+M}$. The relativistic corrections
of order $(\alpha Z)^{4}\frac{m}{M}mc^{2}$ can be found by employing
the Breit equation [1]. A theory of the nuclear recoil effect in higher
orders in $\alpha Z$ must be constructed in the framework of quantum
electrodynamics (QED) on the basis of an exact relativistic equation
 for hydrogen-like atom. Such an equation was proposed by Bethe and
Salpeter [2] immediately after creation of QED. On the basis of this
 equation the nuclear recoil corrections were calculated in [3] up
to terms of order $(\alpha Z)^{5}\frac{m}{M} mc^{2}$.
 It was shown
in this work that the nuclear recoil effect in the case of a complex
nucleus is calculated in a good approximation by assuming the
nucleus is the Dirac particle with the charge $|e|Z$ and the mass M.
Subsequently these corrections were recalculated by a number of
authors [4-6]. Calculations of the nuclear recoil corrections of
the next order in $\alpha Z$ were considered in [7-11].

In the theory of high-Z one-electron ions the parameter $\alpha Z$
 can no longer be considered small. For this reason calculations
 of the nuclear recoil corrections for such systems must be carried
 out without expansion in $\alpha Z$.
In contrast to other QED effects in the region of strongly
bound states ($\alpha Z \sim 1$), the calculation of the nuclear
recoil effect at high $Z$ demands using QED
 outside the external field approximation. ( Calculations of QED
effects  in hydrogen,
positronium, and muonium correspond to the case of weakly
bound states ($\alpha Z \ll 1$).)
 In this connection a non-trivial
problem of derivation of closed expressions for the nuclear recoil
corrections in all orders in $\alpha Z$ arises.
This problem was first discussed in [12,13]. The work [12] was based on the
Bethe-Salpeter equation. This approach encountered serious technical
difficulties, associated with summation of a complete sequence
of irreducible diagrams. These difficulties were partly overcome
only in the lowest orders in $\alpha Z$. Complete $\alpha Z$-dependence
expressions were not found in this way. In [13] a general case of a
relativistic few-electron atom was considered. An efficient method
for summing of the Feynman diagrams in the zeroth and first orders
in $\frac{m}{M}$, based on an expansion of the nuclear propagator,
was proposed in this paper. However, because the procedure of
 the derivation of
the nuclear recoil corrections was not rigorously formulated, the
method considered  there gave several ambiguities in the
expressions for the nuclear recoil corrections. In addition,
certain errors were made in derivation of the formulas for
the contributions with one and two transverse
photons. As result, only a part of the expressions for the relativistic
nuclear recoil corrections was found in this work.  The complete
expressions for the nuclear recoil corrections for hydrogen-like
atoms were obtained in [14]
(the overall sign of  the
 two-transverse-photons contribution was corrected in [15,16]).
The paper [14] was based on a version of the quasipotential approach
that immediately gives the Dirac equation in the limit of infinite
nuclear mass [17,18,5].
( The quasipotential approach
was first introduced in quantum field theory by Logunov and Tavkhelidze
[19] and was subsequently developed by many authors (see, e.g., [20]).
 This approach is absolutely rigorous and, in contrast to the
 Bethe-Salpeter equation, allows one to exclude the relative time
(energy) in the wavefunction from the very beginning. The quasipotential
equation can be represented in the evidently covariant form [20,17].)
 The relevant quasipotential equation
 in the center-of-mass system
is (the relativistic units $\hbar=c=1$ are used)
\begin{eqnarray}
(E-\sqrt{{\bf p}^{2}+M^{2}}-\mbox{\boldmath $\alpha$}{\bf p}-
\beta m)\psi({\bf p})=\int V(E,{\bf p},{\bf q})\psi({\bf q})d{\bf q}\,,
\end{eqnarray}
where $\mbox{\boldmath $\alpha$}$, $\beta$ are the Dirac matrices
acting on the electron variables. The quasipotential $V(E)$ can be
constructed by various methods [17,19,20]. One of the methods consists
in using the relativistic scattering amplitude with one particle
 (nucleus) on
mass shell [17,18,21]. In this method the quasipotential $V(E)$ may
be defined by the Lippman-Schwinger equation
\begin{eqnarray}
V=\tau(1+F\tau)^{-1}\,,
\end{eqnarray}
where
\begin{eqnarray}
F&=&[E-(\sqrt{{\bf p}^{2}+M^{2}}+\mbox{\boldmath $\alpha$}{\bf p}+
\beta m)(1-i0)]^{-1}\,,\\
\tau(E,{\bf p},{\bf q})&=&-2\pi i\beta \overline{u}(-{\bf p})
 T(p_{1},p_{2};q_{1},q_{2})u(-{\bf q})\,,\\
{\bf p}_{1}&=&-{\bf p}_{2}\equiv {\bf p}\,,\;\;\;\;\;\;
{\bf q}_{1}=-{\bf q}_{2}\equiv {\bf q}\,,\nonumber\\
p_{1}^{0}&=&E-\sqrt{{\bf p}^{2}+M^{2}}\,,\;\;\;\;\;\;\;
p_{2}^{0}=\sqrt{{\bf p}^{2}+M^{2}}\,,\nonumber\\
q_{1}^{0}&=&E-\sqrt{{\bf q}^{2}+M^{2}}\,,\;\;\;\;\;\;\;
q_{2}^{0}=\sqrt{{\bf q}^{2}+M^{2}}\,,\nonumber
\end{eqnarray}
$p_{1},\,q_{1}$ are the electron variables, $ p_{2},\,q_{2}$
are the nucleus variables; $T$ is the off-mass-shell relativistic
scattering amplitude; $u({\bf q})$ is the wavefunction of the free
nucleus with the positive energy normalized
by the condition $u^{\dag}({\bf q})u({\bf q})=1\,$.
In [14]  the quasipotential $V(E)$ was constructed
in the zeroth and first orders in $\frac{m}{M}$.
So, the closed expressions for the nuclear recoil corrections
in the first order in $\frac{m}{M}$ and in all orders in $\alpha Z$
were obtained. The most detailed derivation was published in [22].
In [16] these results were generalized to the case of high Z few-electron
atoms. For that a more general method was developed.
 In the second section of the present paper
we briefly formulate the results  of [16].
 In the third section the calculation of the nuclear recoil
corrections for hydrogen-like atoms is considered. In the fourth
section the corrections for high $Z$ lithium-like atoms are
calculated.
%%%%%%%%%%%%%%%%%%%%%%%%%%%%%%%%%%%%%%%%%%%%%%%%%%%%%%%%%%%%%%%%%%%%%%%%
\section{Basic formulas}
 We consider the system of Dirac particles: a nucleus with mass
$M$ and $N$ electrons with mass $m$. Following to ideas of the
quasipotential approach  we introduce in the center-of-mass
system the two-time Green function with the nucleus on the mass
shell
\begin{eqnarray}
\lefteqn{G(t',t,{\bf p}',{\bf x}_{1}',...{\bf x}_{N}',{\bf p}',
{\bf x}_{1},...{\bf x}_{N})}\nonumber\\
& &=\langle {\bf p}',\lambda|
T \psi (t',{\bf x}_{1}')\cdots \psi (t',{\bf x}_{N}')
 \psi^{\dag} (t,{\bf x}_{N})\cdots \psi^{\dag} (t,{\bf x}_{1})
|{\bf p},\lambda\rangle\,,
\end{eqnarray}
where $\psi (x)$ is the electron-positron field operator in the
Heisenberg representation, $T $ is the time ordered product operator;
\begin{eqnarray}
|{\bf p},\lambda \rangle=a_{in}({\bf p},\lambda)|0\rangle\,,\;\;\;\;\;
|{\bf p}',\lambda \rangle=a_{out}({\bf p}',\lambda)|0\rangle\,
\end{eqnarray}
are the {\it in} and {\it out} states of the nucleus; ${\bf p}$ and
$\lambda$ are  momentum and  polarization of the nucleus.
Here we normalize  the operators $a_{in}$ and $a_{out}$ by
\begin{eqnarray}
\{a_{in}^{\dag}({\bf p},\lambda),\,a_{in}({\bf p}',\lambda ')\}=
\{a_{out}^{\dag}({\bf p},\lambda),\,a_{out}({\bf p}',\lambda ')\}=
\delta_{\lambda \lambda '}\delta ({\bf p}-{\bf p}')\,.
\end{eqnarray}
Let us introduce the Fourier transform of $G$:
\begin{eqnarray}
\lefteqn{\delta(E-E')\delta({\bf P}-{\bf P}')
\overline{G}(E,{\bf p}',{\bf p}_{1}',...{\bf p}_{N}',{\bf p},
{\bf p}_{1},...{\bf p}_{N})}\nonumber\\
& &=\frac{1}{2\pi i}\frac{1}{N!}\frac{1}{(2\pi)^{3N}}\int\,
dtdt'd{\bf x}_{1}\cdots d{\bf x}_{N}d{\bf x}_{1}'\cdots d{\bf x}_{N}'
\exp{[i({\cal E}'t'-{\cal E}t)]}\nonumber\\
& &\times \exp{[-i\sum_{i=1}^{N}({\bf p}_{i}'{\bf x}_{i}'-
{\bf p}_{i}{\bf x}_{i})]}G(t',t,{\bf p}',{\bf x}_{1}',...{\bf x}_{N}',{\bf p}',
{\bf x}_{1},...{\bf x}_{N})\,,
\end{eqnarray}
where
\begin{eqnarray}
E={\cal E}+\sqrt{{\bf p}^{2}+M^{2}}-M\,,\;\;\;\;
E'={\cal E}'+\sqrt{{\bf p}'^{2}+M^{2}}-M\,,\nonumber\\
{\bf P}={\bf p}+\sum_{i=1}^{N}{\bf p}_{i}\,,\;\;\;\;
{\bf P}'={\bf p}'+\sum_{i=1}^{N}{\bf p}_{i}'\,.
\end{eqnarray}
In the center-of-mass system we have
$$
{\bf p}=-\sum_{i=1}^{N}{\bf p}_{i}\,,\;\;\;\;
{\bf p}'=-\sum_{i=1}^{N}{\bf p}_{i}'\,.
$$
Let we are interested in the energy of a bound state $n$ of the atom.
The spectral representation of $\overline{G}(E)$ gives
\begin{eqnarray}
\overline{G}(E)=\frac{\Phi_{n}\Phi_{n}^{\dag}}{E-E_{n}}
 + \mbox{ terms regular
at } E= E_{n}\,,
\end{eqnarray}
where $E_{n}$ is the bound state energy with the nucleus rest mass
subtracted, the wavefunction $\Phi_{n}$ is defined by equation
\begin{eqnarray}
\lefteqn{(2\pi)^{\frac{3}{2}}\delta({\bf P})\Phi_{n}({\bf p},
{\bf p}_{1},...{\bf p}_{N})}\nonumber\\
& &=\frac{1}{\sqrt{N!}}\frac{1}{(2\pi)^{\frac{3N}{2}}}
\int d{\bf x}_{1}\cdots d{\bf x}_{N}\exp{[-\sum_{i=1}^{N}
{\bf p}_{i}{\bf x}_{i}]}\nonumber\\
& &\times\langle {\bf p}|
 \psi (0,{\bf x}_{1})\cdots \psi (0,{\bf x}_{N})|n\rangle\,.
\end{eqnarray}
The Green function $\overline{G}(E)$ is constructed by
perturbation theory after  transition in (5) to the
interaction representation.
Let the energy level $n$ belong to a $m$-fold degenerate
level $E_{n}^{(0)}$ in the limit $M\rightarrow \infty$ if
the radiative and interelectronic interaction
 corrections are neglected. (The neglect of the interelectronic
interaction in the zeroth approximation is justified
for high Z few-electron atoms ($N\ll Z$).)
The $m$-dimensional subspace generated by the unperturbed
eigenstates making up this level we designate as $\Omega$.
The projector on $\Omega$ is
\begin{eqnarray}
P_{0}=\sum_{k=1}^{m}u_{k}u_{k}^{\dag}\,,
\end{eqnarray}
where
\begin{eqnarray}
u_{k}=\frac{1}{\sqrt{N!}}\sum_{P}(-1)^{P}\psi_{k_{1}}(P1)\cdots
\psi_{k_{N}}(PN)\,,
\end{eqnarray}
$\psi_{k}$ are solutions of the Dirac equation in the Coulomb
field of the nucleus:
\begin{eqnarray}
H\psi_{k}&=&\varepsilon_{k} \psi_{k}\,,\;\;\;\;
H=\mbox{\boldmath $\alpha$}{\bf p}+\beta m+V_{c}\,,\nonumber\\
E_{n}^{(0)}&=&\sum_{i=1}^{N}\varepsilon_{k_{i}}\,.
\end{eqnarray}
Let us introduce the Green function $g$:
\begin{eqnarray}
g=P_{0}\overline{G}P_{0}\,.
\end{eqnarray}
For this Green function, like $\overline{G}(E)$, we have
\begin{eqnarray}
g(E)=\frac{\phi_{n}\phi_{n}^{\dag}}{E-E_{n}}
 + \mbox{ terms regular
at } E= E_{n}\,,
\end{eqnarray}
where $\phi_{n}=P_{0}\Phi_{n}$ belongs to the subspace $\Omega$.
Constructing $g(E)$ by the perturbation theory in the interaction
representation we get it in the form of a series in powers in $\alpha Z$.
However, we are interested in an expansion in another parameter,
namely, $\frac{m}{M}$. For this reason it is necessary to sum
infinite sequences of the Feynman diagrams in the zeroth and first
orders in $\frac{m}{M}$. We designate the contribution of the terms of
the zeroth order in $\frac{m}{M}$ by $g_{0}$. In [16] it was found
\begin{eqnarray}
g_{0}(E)=\frac{P_{0}}{E-E_{n}^{(0)}}\,.
\end{eqnarray}
 From the equation (16) and the identity
\begin{eqnarray}
g^{-1}g=1
\end{eqnarray}
we obtain for $E=E_{n}$
\begin{eqnarray}
g^{-1}(E_{n})\phi_{n}=0\,.
\end{eqnarray}
Or, introducing the quasipotential operator
\begin{eqnarray}
v(E)=g_{0}^{-1}-g^{-1}=g_{0}^{-1}\,\Delta g\,g_{0}^{-1}
+g_{0}^{-1}\,\Delta g\,g_{0}^{-1}\,\Delta g\,g_{0}^{-1}+\cdots\,,
\end{eqnarray}
where $\Delta g\equiv g-g_{0}\,$, we obtain
\begin{eqnarray}
(E_{n}^{(0)}+v(E_{n}))\phi_{n}=E_{n}\phi_{n}\,.
\end{eqnarray}
It follows the equation for determination of the energy levels
\begin{eqnarray}
{\rm det}\{(E-E_{n}^{(0)})\delta_{ik}-v_{ik}(E)\}=0\,.
\end{eqnarray}
It should be stressed that equation (22) is absolutely rigorous
within QED and  gives, in principle, the exact energies of the
$m$ levels arising from the $m$-fold degenerate level $E_{n}^{(0)}$.
In [16] the quasipotential $v_{ik}$ was found in the first
order in $\frac{m}{M}$ and in the zeroth order in $\alpha$
(but in all orders in $\alpha Z$) by summing
 infinite sequences of the Feynman diagrams in the
Coulomb gauge. For that the expansion of the nuclear propagator
from [13] was used.
Only the following kinds of the diagrams contribute in the
considered order:
\begin{itemize}
\item The diagrams with only  Coulomb photons.
\item The diagrams with one  transverse and arbitrary number
of Coulomb photons.
\item The diagrams with two transverse  and arbitrary number
of Coulomb photons.
\end{itemize}
The contribution  from the diagrams
with only Coulomb photons is
\begin{eqnarray}
(v_{c})_{ik}&=&(v_{c}^{(1)})_{ik}+(v_{c}^{(2)})_{ik}+(v_{c}^{(int)})_{ik}\,,\\
(v_{c}^{(1)})_{ik}&=&\sum_{s=1}^{N}\delta_{i_{1}k_{1}}\cdots
\stackrel{s}{\sqcap}\cdots
\delta_{i_{N}k_{N}}
\langle i_{s}|\frac{{\bf p}_{s}^{2}}{2M}|k_{s}\rangle\,,\\
(v_{c}^{(2)})_{ik}&=&
\frac{2\pi i}{M}\sum_{s=1}^{N}\delta_{i_{1}k_{1}}\cdots
\stackrel{s}{\sqcap}\cdots
\delta_{i_{N}k_{N}}
\int_{-\infty}^{\infty}d\omega\,
\delta_{+}^{2}(\omega)\nonumber\\
&&\times\langle i_{s}|
[{\bf p}_{s},v_{s}]G_{s}(\omega+\varepsilon_{i_{s}})
[{\bf p}_{s},v_{s}]|k_{s}\rangle\,,\\
(v_{c}^{(int)})_{ik}&=&\frac{1}{M}\sum_{s<s'}\delta_{i_{1}k_{1}}\cdots
\stackrel{s}{\sqcap}\cdots\stackrel{s'}{\sqcap}\cdots
\delta_{i_{N}k_{N}}\nonumber\\
&&\times\sum_{P}(-1)^{P}\langle Pi_{s}Pi_{s'}|
{\bf p}_{s}{\bf p}_{s'}|k_{s}k_{s'}\rangle\,,
\end{eqnarray}
where $|i_{s}\rangle$ and $|k_{s}\rangle$
 are the one-electron unperturbed states of the Dirac electron
in the Coulomb field of the nucleus, belonging to the N-electron
states $i$ and $k$, respectively;
${\bf p}$ is the momentum operator,
$v_{s}\equiv V_{c}(r_{s})=-\frac{\alpha Z}{r_{s}}\,$;
the symbol $\stackrel{s}{\sqcap}$ means that the factor $\delta_{i_{s}k_{s}}$
is omitted in the product;
 $\delta_{+}(\omega)=\frac{i}{2\pi}
(\omega+i0)^{-1}$, $G(\omega)=(\omega-H(1-i0))^{-1}$ is the relativistic
Coulomb Green function.
( Formally, the matrix element in equation (25)
at fixed $\omega$ is infinite, due to the strong Coulomb singularity
at $r=0\,$. It means that the integration over $\omega$ must be carried
out on an intermediate stage of the calculation, depending on which
representation of $G$ is used.)
 The contribution from the diagrams with one transverse
and arbitrary number of Coulomb photons consists of two terms. The
first term depends on the spin of the nucleus and coincides with
the Fermi-Breit expression for the hyperfine interaction [23].
The second term is
\begin{eqnarray}
(v_{tr(1)})_{ik}&=&
(v_{tr(1)}^{(1)})_{ik}+(v_{tr(1)}^{(2)})_{ik}+(v_{tr(1)}^{(int)})_{ik}\,,\\
(v_{tr(1)}^{(1)})_{ik}&=&-\frac{1}{2M}
\sum_{s=1}^{N}\delta_{i_{1}k_{1}}\cdots
\stackrel{s}{\sqcap}\cdots
\delta_{i_{N}k_{N}}\nonumber\\
&&\times\langle i_{s}| \Bigl({\bf D}_{s}(0){\bf p}_{s}+
{\bf p}_{s}{\bf D}_{s}(0)\Bigr) |k_{s}\rangle\,,\\
(v_{tr(1)}^{(2)})_{ik}&=&
-\frac{1}{M}\sum_{s=1}^{N}\delta_{i_{1}k_{1}}\cdots
\stackrel{s}{\sqcap}\cdots
\delta_{i_{N}k_{N}}
\int_{-\infty}^{\infty}d\omega\,
\delta_{+}(\omega)\nonumber\\
&&\times\langle i_{s}|\Bigl(
[{\bf p}_{s},v_{s}]G_{s}
(\omega+\varepsilon_{i_{s}}){\bf D}_{s}(\omega)\nonumber\\
&&-{\bf D}_{s}(\omega)G_{s}(\omega+\varepsilon_{i_{s}})
[{\bf p}_{s},v_{s}]\Bigr)|k_{s}\rangle\,,\\
(v_{tr(1)}^{(int)})_{ik}&=&-\frac{1}
{M}\sum_{s<s'}\delta_{i_{1}k_{1}}\cdots
\stackrel{s}{\sqcap}\cdots\stackrel{s'}{\sqcap}\cdots
\delta_{i_{N}k_{N}}\nonumber\\
&&\times\sum_{P}(-1)^{P}\langle Pi_{s}Pi_{s'}|\Bigl(
{\bf D}_{s}(\varepsilon_{Pi_{s}}-\varepsilon_{k_{s}})
{\bf p}_{s'}\nonumber\\
&&+ {\bf p}_{s}
{\bf D}_{s'}(\varepsilon_{Pi_{s'}}-\varepsilon_{k_{s'}})
 \Bigr)|k_{s}k_{s'}\rangle\,,
\end{eqnarray}
where
\begin{eqnarray}
D_{m}(\omega)=-4\pi\alpha Z\alpha_{l}D_{lm}(\omega)\,,
\end{eqnarray}
$\alpha_{l}\;(l=1,2,3)$ are the Dirac matrices, $ D_{lm}(\omega)$ is
the transverse part of the photon propagator in the Coulomb gauge.
In the coordinate representation it is
\begin{eqnarray}
D_{ik}(\omega,{\bf r})=-\frac{1}{4\pi}\Bigl\{\frac
{\exp{(i|\omega|r)}}{r}\delta_{ik}+\nabla_{i}\nabla_{k}
\frac{(\exp{(i|\omega|r)}
-1)}{\omega^{2}r}\Bigr\}\,.
\end{eqnarray}
The contribution from the diagrams with two transverse
and arbitrary number of Coulomb photons is
\begin{eqnarray}
(v_{tr(2)})_{ik}&=&
(v_{tr(2)}^{(1)})_{ik}+(v_{tr(2)}^{(int)})_{ik}\,,\\
(v_{tr(2)}^{(1)})_{ik}&=&\frac{i}{2\pi M}
\sum_{s=1}^{N}\delta_{i_{1}k_{1}}\cdots
\stackrel{s}{\sqcap}\cdots
\delta_{i_{N}k_{N}}\nonumber\\
&&\times \int_{-\infty}^{\infty}d\omega\langle i_{s}|
{\bf D}_{s}(\omega)G_{s}(\omega +\varepsilon_{i_{s}}){\bf D}_{s}
(\omega)|k_{s}\rangle\,,\\
(v_{tr(2)}^{(int)})_{ik}&=&\frac{1}
{M}\sum_{s<s'}\delta_{i_{1}k_{1}}\cdots
\stackrel{s}{\sqcap}\cdots\stackrel{s'}{\sqcap}\cdots
\delta_{i_{N}k_{N}}\nonumber\\
&&\times\sum_{P}(-1)^{P}\langle Pi_{s}Pi_{s'}|
{\bf D}_{s}(\varepsilon_{Pi_{s}}-\varepsilon_{k_{s}})\nonumber\\
&&\times{\bf D}_{s'}(\varepsilon_{Pi_{s'}}-\varepsilon_{k_{s'}})
 |k_{s}k_{s'}\rangle\,.
\end{eqnarray}

The formulas (23)-(35) were derived in [16].
The corresponding formulas for the case of a one-electron atom
were first obtained in [14] (the overall sign
of the contribution $\Delta E_{tr(2)}$ was corrected in [15,16])
and recently reproduced in [24,10].

The contributions $v_{c}^{(1)}$, $v_{c}^{(int)}$,
 $v_{tr(1)}^{(1)}$, and $v_{tr(1)}^{(int)}$ are leading
for low $\alpha Z$ and  completely define
the nuclear recoil corrections
within  $\frac{m^{2}}{M}(\alpha Z)^{4}$ approximation.
It follows that within $\frac{m^{2}}{M}(\alpha Z)^{4}$ approximation
the nuclear recoil corrections can be obtained by evaluating the
expectation values with the Dirac wavefunctions of the operator
\begin{eqnarray}
H_{M}=\frac{1}{2M}\sum_{s,s'}\Bigl({\bf p}_{s}{\bf p}_{s'}-
\frac{\alpha Z}{r_{s}}\Bigl(\mbox{\boldmath $\alpha$}_{s}+\frac
{(\mbox{\boldmath $\alpha$}_{s}{\bf r}_{s}){\bf r}_{s}}{r_{s}^{2}}
\Bigr){\bf p}_{s'}\Bigr)
\end{eqnarray}
In [25] the relativistic nuclear recoil corrections
of order  $\frac{m^{2}}{M}(\alpha Z)^{4}$ to the energy levels of two-
 and three-electron multicharged ions were calculated using
this operator.
The expression (36) can  be found by reformulating the
Stone's theory as well [26].

%%%%%%%%%%%%%%%%%%%%%%%%%%%%%%%%%%%%%%%%%%%%%%%%%%%%%%%%%%%%%%%%%%%%%%%%%%%
\section{Hydrogen-like atoms}
For hydrogen-like atoms the nuclear recoil corrections
to the energy of a state $a$ are defined by the diagonal matrix elements
($\Delta E =(v)_{aa}$) of the one-electrons contributions (24),(25),(28)
,(29), and (34).
The terms $\Delta E_{c}^{(1)}$ and $\Delta E_{tr(1)}^{(1)}$ are leading
at low $Z$. These terms can easily be calculated by using
the virial relations for the Dirac  equation [27-29]. Such a calculation
 gives [14]
\begin{eqnarray}
\Delta E_{c}^{(1)}&=&\frac{m^{2}}{2M}\Bigl\{1-\frac{(\gamma+n_{r})^{2}}
{N^{2}}+\frac{2(\alpha Z)^{4}}{N^{4}\gamma(4\gamma^{2}-1)}
[\kappa(2\kappa(\gamma+n_{r})-N)\nonumber\\
&&+n_{r}(4\gamma^{2}-1)]\Bigl\}\,,\\
\Delta E_{tr(1)}^{(1)}&=&-\frac{m^{2}}{M}\frac{(\alpha Z)^{4}}
{N^{4}\gamma(4\gamma^{2}-1)}
[\kappa(2\kappa(\gamma+n_{r})-N)
+n_{r}(4\gamma^{2}-1)]\,,\\
\Delta E^{(1)}&\equiv&\Delta E_{c}^{(1)}+\Delta E_{tr(1)}^{(1)}=
\frac{m^{2}-\varepsilon_{a}^{2}}{2M}=\frac{m^{2}}{M}\frac{(\alpha Z)^{2}}
{2N^{2}}\,,
\end{eqnarray}
where
$$
\kappa=(-1)^{j+l+\frac{1}{2}}(j+\frac{1}{2})\;,\;\;\;\;\;\;\;
\gamma=\sqrt{\kappa^{2}-(\alpha Z)^{2}}\,,
$$
$$
N=\sqrt{n^{2}-2n_{r}(|\kappa|-\gamma)}\;,\;\;\;\;\;n=n_{r}+|\kappa|\,,
$$
$j$ is the total electron moment, $l$ is the orbital moment,
$n$ is the principal quantum number,
$n_{r}$ is the radial quantum number.
Only these terms contribute within the $\frac{m^{2}}{M}(\alpha Z)^{4}$
approximation. Expanding (39) in power series in $\alpha Z$
we find
\begin{eqnarray}
\Delta E^{(1)}&=&\frac{m^{2}}{M}\Bigl\{\frac{(\alpha Z)^{2}}{2n^{2}}+
\frac{(\alpha Z)^{4}}{2n^{3}}\Bigl(\frac{1}{j+\frac{1}{2}}-\frac{1}{n}
\Bigr)\nonumber\\
&&+\frac{(\alpha Z)^{6}n_{r}}{2n^{4}(j+\frac{1}{2})^{2}}\Bigl(\frac{1}{4(j
+\frac{1}{2})}
+\frac{n_{r}}{n^{2}}\Bigr)+\cdots\Bigr\}\,,
\end{eqnarray}

The terms $\Delta E_{c}^{(2)},\;\Delta E_{tr(1)}^{(2)},$ and
$\Delta E_{tr(2)}^{(1)}$ ( the equations (25),(29), and (34) )
 are given in the form that allows one to use the
relativistic Coulomb Green function for their calculations.
In addition, this form is convenient for $\alpha Z$-expansion
calculations [10].
However, in the present paper
 we  transform these equations
 to ones that are most convenient for calculations  using
 the finite basis set methods [30-32].

Integrating over $\omega$ in (25) we find
\begin{eqnarray}
\Delta E_{c}^{(2)}=-\frac{1}{M}\sum_{\varepsilon_{n}<0}\langle a|{\bf p}|
n\rangle\langle n|{\bf p}|a\rangle\,.
\end{eqnarray}
(It should be noted here that the formula (41) was first found in [13].
Its derivation was refined in [14]. A similar formula but with the
projector on the negative energy states of a free electron was obtained
in the lowest order in $\alpha Z$ in [12].)
The matrix elements of the momentum operator are easily calculated
 using the identity [25]
\begin{eqnarray}
{\bf p}=\frac{1}{2}(\mbox{\boldmath $\alpha$}H+H\mbox{\boldmath $\alpha$})
- \mbox{\boldmath $\alpha$}V_{c}\,.
\end{eqnarray}

Rotating in (29) the integration contour in the complex $\omega$ plane
 we find
\begin{eqnarray}
\Delta E_{tr(1)}^{(2)}&=&\Delta E_{tr(1)}^{(2,a)}+\Delta E_{tr(1)}^{(2,b)}
+ \Delta E_{tr(1)}^{(2,c)}\,,\\
\Delta E_{tr(1)}^{(2,a)}&=&\frac{1}{2M}\sum_{\varepsilon_{n}\not=
\varepsilon_{a}}
\Bigl\{\langle a|
{\bf p}|n\rangle\langle n|
{\bf D}(0)|a\rangle\nonumber\\
&&+\langle a|{\bf D}(0)|n\rangle\langle n|{\bf p}|a\rangle\Bigr\}\,,\\
\Delta E_{tr(1)}^{(2,b)}&=&\frac{2}{\pi M}{\rm Re}\int_{0}^{\infty}dy\sum_{
\varepsilon_{n}\not=
\varepsilon_{a}}
\frac{\varepsilon_{a}-\varepsilon_{n}}
{y^{2}+(\varepsilon_{a}-\varepsilon_{n})^{2}}
\langle a|{\bf p}|n\rangle\langle n|
{\bf S}(y)|a\rangle\,,\\
\Delta E_{tr(1)}^{(2,c)}&=&-\frac{1}{M}\sum_{|\varepsilon_{n}|<
\varepsilon_{a}}
\Bigl\{\langle a|
{\bf p}|n\rangle\langle n|
{\bf D}(\varepsilon_{a}-\varepsilon_{n})|a\rangle\nonumber\\
&&+\langle a|{\bf D}(\varepsilon_{a}-\varepsilon_{n})
|n\rangle\langle n|{\bf p}|a\rangle\Bigr\}\,,
\end{eqnarray}
where
$$
{\bf S}(y)={\bf S}_{1}(y)+{\bf S}_{2}(y)\,,
$$
$$
{\bf S}_{1}(y)=\alpha Z \mbox{\boldmath $\alpha$}\frac{\exp{(-yr)}}{r}\,,
$$
$$
{\bf S}_{2}(y)=i\alpha Z[H,\tilde f(y,r){\bf n}]\,,
$$
$$
\tilde f(y,r)=\frac{\exp{(-yr)}(1+yr)-1}{y^{2}r^{2}}\,,
$$
$$
{\bf D}(\omega)={\bf D}_{1}(\omega)+{\bf D}_{2}(\omega)\,,
$$
$$
{\bf D}_{1}(\omega)=\alpha Z \mbox{\boldmath $\alpha$}
\frac{\exp{(i|\omega|r)}}{r}\,,
$$
$$
{\bf D}_{2}(\omega)=i\alpha Z[H, f(\omega,r){\bf n}]\,,
$$
$$
f(\omega,r)=\frac{1-\exp{(i|\omega| r)}(1-i|\omega|r)}{\omega^{2}r^{2}}\,,
$$
$$
{\bf D}(0)=\alpha Z\frac{\mbox{\boldmath $\alpha$}}{r}-\frac{i\alpha Z}{2}
[H,{\bf n}]\,,
$$
${\bf n}=\frac{\bf r}{r}$.
The term $\Delta E_{tr(1)}^{(2,c)}$ has real and imaginary parts.
The imaginary part  gives a small correction to the width
of the level.
Integrating over $y$ in (45) and uniting the contributions
$\Delta E_{tr(1)}^{(2,a)}$,  $\Delta E_{tr(1)}^{(2,b)}$, and the real
part of $\Delta E_{tr(1)}^{(2,c)}$
  we find
\begin{eqnarray}
\Delta E_{tr(1)}^{(2)}&=&\frac{2\alpha Z}{\pi M}{\rm Re}\sum_{
\varepsilon_{n}\not=\varepsilon_{a}}(\varepsilon_{n}
-\varepsilon_{a})\langle a|i \mbox{\boldmath $\alpha$}\phi(r)|n\rangle
\nonumber\\
&&\times \langle n|[i \mbox{\boldmath $\alpha$}\Phi_{1}(r)
+{\bf n}\Phi_{2}(r)]|a\rangle\,,
\end{eqnarray}
where
\begin{eqnarray}
\phi&=&\frac{\varepsilon_{a}+\varepsilon_{n}}{2}+\frac{\alpha Z }{r}\,,\\
\Phi_{1}(r)&=&\frac{1}{\Delta_{n}r}[{\rm ci}(\Delta_{n}r)\,
{\rm sin}(\Delta_{n}r)-
{\rm si}(\Delta_{n}r)\,{\rm cos}(\Delta_{n}r)\nonumber\\
&&+{\rm sign}(\varepsilon_{a}-\varepsilon_{n})\frac{\pi}{2}]
-\theta (\varepsilon_{a}-|\varepsilon_{n}|)\frac{\pi}{\Delta_{n}}
\frac{\exp{(i\Delta_{n}r)}}{r}\,,\\
\Phi_{2}(r)&=&-{\rm sign}
(\varepsilon_{a}-\varepsilon_{n})\frac{1}{(\Delta_{n}r)^{2}}
\{-{\rm si}
(\Delta_{n}r)\,{\rm cos}(\Delta_{n}r)-\frac{\pi}{2}+\Delta_{n}r\nonumber\\
&&+{\rm ci}(\Delta_{n}r)[{\rm sin}
(\Delta_{n}r)-(\Delta_{n}r)\,{\rm cos}(\Delta_{n}r)]\nonumber\\
&&-(\Delta_{n}r)\,{\rm si}(\Delta_{n}r)\,{\rm sin}
(\Delta_{n}r)\}-\frac{\pi}{4}
-\theta(\varepsilon_{a}-|\varepsilon_{n}|)\pi f(\Delta_{n},r)\,,
\end{eqnarray}
$\Delta_{n}=|\varepsilon_{a}-\varepsilon_{n}|$,
$\theta(x)=(x+|x|)/2x\,$.

The contribution $\Delta E_{tr(2)}$   is equal
\begin{eqnarray}
\Delta E_{tr(2)}^{(1)}&=&\Delta E_{tr(2)}^{(1,a)}+\Delta E_{tr(2)}^{(1,b)}
+ \Delta E_{tr(2)}^{(1,c)}\,,\\
\Delta E_{tr(2)}^{(1,a)}&=&-\frac{1}{\pi M}\int_{0}^{\infty}dy
\sum_{\varepsilon_{n}\not=\varepsilon_{a}}
\frac{\varepsilon_{a}-\varepsilon_{n}}
{y^{2}+(\varepsilon_{a}-\varepsilon_{n})^{2}}
\langle a|{\bf S}(y)|n\rangle\langle n|
{\bf S}(y)|a\rangle\,,\\
\Delta E_{tr(2)}^{(1,b)}&=&\frac{1}{2M}\sum_{\varepsilon_{n}=
\varepsilon_{a}}\langle a|{\bf D}(0)|n\rangle \langle n|{\bf D}(0)|a\rangle
\,,\\
 \Delta E_{tr(2)}^{(1,c)}&=&\frac{1}{M}\sum_{|\varepsilon_{n}|<
\varepsilon_{a}}\langle a|{\bf D}(\varepsilon_{a}-\varepsilon_{n}
)|n\rangle \langle n|{\bf D}(\varepsilon_{a}-\varepsilon_{n}
)|a\rangle \,.
\end{eqnarray}
The term $\Delta E_{tr(2)}^{(1,c)}$, like $\Delta E_{tr(1)}^{(2,c)}$,
has an imaginary part which gives a small contribution to the width
of the level.

After integration over angles that is easily carried out using formulas
presented in Appendix, the calculation of the expressions (41),
(47), and (52)-(54)
 was done using
 the B-spline method
for the Dirac equation, developed in [31].
 The zero boundary conditions and the grid selection
algorithm proposed in [33] were used. However,
 we used the grid $r_{i}=
\frac{\rho_{i}^{4}\gamma_{0}}{Z}$,
 where $\gamma_{0}=\sqrt{1-(\alpha Z)^{2}}\,$,
instead of the grid $r_{i}=
\frac{\rho_{i}^{4}}{Z}\,$  [33].
The radial integration caused no problems and was carried out
with high accuracy using the Gauss-Legendre quadratures.
The integration over $y$ in (52) was also done by the
Gauss-Legendre quadratures with a suitable transformation
to map the infinite integration range to a finite one.
The uncertainty of the integration was estimated from the
stability of the result with respect to change of the number
of integration points and the grid parameters and was found
to be much smaller than the uncertainty due to the finiteness
of the basis set. The size of the box was chosen to be
sufficiently large so as not to affect the results.
The uncertainty, due to the finiteness of the basis set, was
 estimated by changing the number of splines from 40 to 90.
In addition, to make an independent estimate of the uncertainty
of the numerical results we calculated the corrections
$ \Delta E_{c}^{(2)}$ and $\Delta E_{tr(1)}^{(2)}$ using two
different representations for them. So,  the correction
$ \Delta E_{c}^{(2)}$ was calculated by the formula (41)
as well as by
\begin{eqnarray}
\Delta E_{c}^{(2)}=-\frac{1}{M}\Bigl\{\langle a|{\bf p}^{2}|a\rangle-
\sum_{\varepsilon_{n}>0}\langle a|{\bf p}|
n\rangle\langle n|{\bf p}|a\rangle\Bigr\}\,.
\end{eqnarray}
We found that the results of  both calculations
coincided with each other with good precision, and this coincidence
 improved when the number of splines increased. The correction
$\Delta E_{tr(1)}^{(2)}$ was calculated by the equation (47)
as well as by (43)-(45). The results of  both calculations
coincided with each other with high accuracy.

Table 1, 2 and 3
 show the results of the numerical calculation
for the $1s$, $2s$, and $2p_{\frac{1}{2}}$ states,
 respectively, expressed in terms of
the function $P(\alpha Z)$ defined by
\begin{eqnarray}
\Delta E^{(2)} = \Delta E_{c}^{(2)}+\Delta E_{tr(1)}^{(2)}+
\Delta E_{tr(2)}^{(1)}=
\frac{m}{M}\frac{(\alpha Z)^{5}}{\pi n^{3}}P(\alpha Z)mc^{2}
\end{eqnarray}
The functions $P_{c}$, $P_{tr(1)}$, and $P_{tr(2)}$  correspond to the
contributions $\Delta E_{c}^{(2)}$,
 $\Delta E_{tr(1)}^{(2)}$, and $\Delta E_{tr(2)}^{(1)}$,
respectively. For comparison, in the last columns of the tables  Salpeter's
contributions [3-6]
\begin{eqnarray}
P_{S}^{(1s)}
(\alpha Z)&=&-\frac{2}{3}\ln{(\alpha Z)}-\frac{8}{3}\,
2.984129+\frac{14}{3}\ln{2}
+\frac{62}{9}\,,\\
P_{S}^{(2s)}(\alpha Z)&=&
-\frac{2}{3}\ln{(\alpha Z)}-\frac{8}{3}\,2.811769
+\frac{187}{18}\,,\\
P_{S}^{(2p_{\frac{1}{2}})}&=&\frac{8}{3}\,0.030017-\frac{7}{18}
\end{eqnarray}
are given. The uncertainties given in the tables  correspond
only  to  errors
of the numerical calculation. In addition, there is an uncertainty due to
deviation from the point single particle model of the nucleus, used here.

To make a more detailed comparison with the $\alpha Z$-expansion calculations
we represent  the functions $P_{c}$, $P_{tr(1)}$, and $P_{tr(2)}$
for the $s$ states
in the form
\begin{eqnarray}
P_{c}&=&a_{1}+a_{2}\alpha Z+a_{3}(\alpha Z)^{2} \ln{(\alpha Z)}+a_{4}
(\alpha Z)^{2}
\,,\nonumber\\
P_{tr(1)}&=&b_{1}\ln{(\alpha Z)}+
b_{2}+b_{3}\alpha Z\ln{(\alpha Z)}\nonumber\\
&&+b_{4}\alpha Z+b_{5}(\alpha Z)^{2} \ln{(\alpha Z)}+b_{6}(\alpha Z)^{2}
+b_{7}(\alpha Z)^{3}\,,\nonumber\\
P_{tr(2)}&=&c_{1}\ln{(\alpha Z)}+
c_{2}+c_{3}\alpha Z\ln{(\alpha Z)}\nonumber\\
&&+c_{4}\alpha Z+c_{5}(\alpha Z)^{2} \ln{(\alpha Z)}+c_{6}(\alpha Z)^{2}
+c_{7}(\alpha Z)^{3}\,.
\end{eqnarray}
The coefficients $a_{i}$, $b_{i}$, and $c_{i}$ can be calculated from our
numerical results for the $P(\alpha Z)$-functions.
 Such a calculation for the $2s$ state
 using the values of the $P(\alpha Z)$-functions for
$Z=$1,2,3,5,8,15,30 gives
\begin{eqnarray}
a_{1}&=&-1.3333\,,\;\;\;\;\;\;a_{2}=3.156\,,\nonumber\\
b_{1}&=&-2.6662\,,\;\;\;b_{2}=-0.091\,,\;\;\;b_{3}=-6.02
\,,\;\;\;b_{4}=-9.98\,,\nonumber\\
 c_{1}&=&2.0031\,,\;\;\;c_{2}=4.338\,,\;\;\;c_{3}=6.46
\,,\;\;\;c_{4}=5.92\,.
\end{eqnarray}
The coefficients $a_{1}$, $b_{1,2}$, and $c_{1,2}$
 are in good agreement with
 Salpeter's results
\begin{eqnarray}
a_{1}&=&-1.3333\,\;\;\;
b_{1}=-2.6666\,,\;\;\;b_{2}=-0.094\,,\;\;\;\nonumber\\
c_{1}&=&2.0000\,,\;\;\;c_{2}=4.318\,.
\end{eqnarray}
Within  errors of the numerical procedure our values
  $b_{3},\;c_{3},$  are  in good agreement with the
analytical result of [8,9]
\begin{eqnarray}
b_{3}=-c_{3}=-2\pi=-6.2832\,,\;\;\; b_{3}+c_{3}=0\,
\end{eqnarray}
(the coefficient $b_{3}$ was first found in [7]).
The coefficient $a_{2}$ coincides, within the numerical errors, with
the corresponding coefficient ($a_{2}=\pi=3.1459$)
obtained in [7]. The coefficients $b_{4}$ and $c_{4}$ are in satisfactory
agreement with the  results of [10]
\begin{eqnarray}
b_{4}=-10.996\,,\;\;\;c_{4}=5.569\,.
\end{eqnarray}
For the $1s$ state we have  found a similar agreement.

 To make  a similar  comparison
 for the $2p_{\frac{1}{2}}$ state
 we represent the functions $P_{tr(1)}$ and
$P_{tr(2)}$ for this state in the form
\begin{eqnarray}
P_{tr(1)}&=&b_{1}+b_{2}\alpha Z+b_{3}(\alpha Z)^{2} \ln{(\alpha Z)}\nonumber\\
&&+b_{4}
(\alpha Z)^{2}+ b_{5}(\alpha Z)^{3} \ln{(\alpha Z)}+b_{6}
(\alpha Z)^{3}+b_{7}(\alpha Z)^{4}\,,\nonumber\\
P_{tr(2)}&=&c_{1}+c_{2}\alpha Z+c_{3}(\alpha Z)^{2} \ln{(\alpha Z)}\nonumber\\
&&+c_{4}
(\alpha Z)^{2}+c_{5}(\alpha Z)^{3} \ln{(\alpha Z)}+c_{6}
(\alpha Z)^{3}+c_{7}(\alpha Z)^{4}\,.
\end{eqnarray}
Using our values of $P(\alpha Z)$ for $Z=1,2,3,5,8,15,30$ we have found
\begin{eqnarray}
b_{1}&=&-0.142178\,,\;\;\;b_{2}=-0.26166\,,\nonumber\\
c_{1}&=&-0.166666\,,\;\;\;c_{2}=1.30881\,.
\end{eqnarray}
The coefficients  $b_{1}$ and $c_{1}$
 are in excellent agreement with
the Salpeter's results:
 $b_{1}=-0.142178$ and $c_{1}=-0.166667$.
Adding to
the sum $b_{2}+c_{2}$ the corresponding coefficient from the
equation (40), we find that the total coefficient of the $\frac
{m^{2}}{M}\frac{(\alpha Z)^{6}}{n^{3}\pi}$ contribution for the
$2p_{\frac{1}{2}}$ state is $ 1.43985$. The related
analytical result obtained in [11] is $\frac{11}{24}\pi= 1.43990$.

 The term $\Delta E^{(1)}$ does not
contribute to the Lamb shift of hydrogen-like atoms.
  The contribution of the difference between   $\Delta E^{(2)}$ and
the Salpeter's correction  to the Lamb shift ($n=2$)
of hydrogen is $-1.32(6)\,kHz$.
The corresponding result for the ground state is $-7.1(9)\,kHz$.
 These results are in good agreement
with analytical calculations of the $\frac{m^{2}}{M}(\alpha Z)^{6}$
contributions [10,11]. So, according to [10] the total
 $\frac{m^{2}}{M}(\alpha Z)^{6}$ correction, including the
related term from the equation (40), is  $-7.4\,kHz$ and $ -0.77\,kHz$
for the $1s$ and $2s$ states, respectively. The
$\frac{m^{2}}{M}(\alpha Z)^{6}$ correction for the $2p_{\frac{1}{2}}$
state, found in [11], is $0.58\,kHz$. (We note that in [11]
the correction of order $\frac{m^{2}}{M}(\alpha)^{2}(\alpha Z)^{4}$
for  $p$ states is also calculated.)

Let us consider the nuclear recoil corrections
for hydrogen-like uranium. According to the formula (39)
the first correction  is
\begin{eqnarray}
\Delta E_{1s}^{(1)}=0.26\,eV\,,\;\;\;\;\;
\Delta E_{2s}^{(1)}=\Delta E_{2p_{\frac{1}{2}}}^{(1)}=0.08\,eV\,.
\end{eqnarray}
The second correction defined by (56) is
\begin{eqnarray}
\Delta E_{1s}^{(2)}=0.24\,eV\,,\;\;\;\;
\Delta E_{2s}^{(2)}=0.05\,eV\,,\;\;\;\;
\Delta E_{2p_{\frac{1}{2}}}^{(2)}=0.01\,eV\,.
\end{eqnarray}
In the next section we use these results to find the total nuclear
recoil contribution to the energy of the $2p_{\frac{1}{2}}-2s$ transition
in lithium-like uranium.
%%%%%%%%%%%%%%%%%%%%%%%%%%%%%%%%%%%%%%%%%%%%%%%%%%%%%%%%%%%%%%%%%%%%%%%%%%%%
\section{High Z lithium-like atoms}
The wavefunction of a high Z lithium-like atom with one electron
over the closed $(1s)^{2}$ shell in the zeroth approximation is
\begin{eqnarray}
u=\frac{1}{\sqrt{3!}}\sum_{P}(-1)^{P}\psi_{1s\uparrow}(P1)
\psi_{1s\downarrow}(P2)\psi_{a}(P3)\,.
\end{eqnarray}
The nuclear recoil correction for the lithium-like atom is the sum of
the one- and two-electron corrections. The one-electron correction
is obtained by summing all the one-electron contributions considered
in the preceeding section over all the one-electron states that are
occupied.
According to (26), (30), and (35) the two-electron corrections for the state
considered here are
\begin{eqnarray}
\Delta E_{c}^{(int)}&=&-\frac{1}{M}\sum_{\varepsilon_{n}=\varepsilon_{1s}}
\langle a|{\bf p}|n\rangle \langle n|{\bf p}|a\rangle\,,\\
\Delta E_{tr(1)}^{(int)}&=&\frac{1}{M}\sum_{\varepsilon_{n}=
\varepsilon_{1s}}
\Bigl\{\langle a|
{\bf p}|n\rangle\langle n|
{\bf D}(\varepsilon_{a}-\varepsilon_{n})|a\rangle\\
&&+\langle a|{\bf D}(\varepsilon_{a}-\varepsilon_{n})
|n\rangle\langle n|{\bf p}|a\rangle\Bigr\}\,,\nonumber\\
 \Delta E_{tr(2)}^{(int)}&=&-\frac{1}{M}\sum_{\varepsilon_{n}
=\varepsilon_{1s}}\langle a|{\bf D}(\varepsilon_{a}-\varepsilon_{n}
)|n\rangle \langle n|{\bf D}(\varepsilon_{a}-\varepsilon_{n}
)|a\rangle \,.
\end{eqnarray}
The terms $\Delta E_{tr(1)}^{(int)}$ and $\Delta E_{tr(2)}^{(int)}$ have
real and imagine parts and are
 cancelled by a part of the one-electron
terms  $\Delta E_{tr(1)}^{(2,c)}$ and $\Delta E_{tr(2)}^{(1,c)}$,
 which corresponds to the $1s$ states. So, for the $(1s)^{2}2s$
and $(1s)^{2}2p_{\frac{1}{2}}$ states the imagine parts of the one- and
two-electron contributions are completely cancelled.

We note here that the nuclear recoil corrections for a high $Z$
lithium-like atom with one electron over the closed $(1s)^{2}$
shell can be
obtained from the nuclear recoil corrections for the hydrogen-like
atom by changing the sign of $i0$ in the denominators of the
electron propagator in the Coulomb field of the nucleus,
corresponding to the states of the closed shell.
It follows, in particular, the sum of the one- and
two-electron  Coulomb contributions
can be represented in a simple form
\begin{eqnarray}
\Delta E_{c}=\frac{1}{2M}\Bigl\{
\sum_{\varepsilon_{n}>\varepsilon_{1s}}|\langle a|{\bf p}|n
\rangle |^{2}-\sum_{\varepsilon_{n}\leq\varepsilon_{1s}}|\langle a|{\bf p}|n
\rangle |^{2}\Bigr\}\,.
\end{eqnarray}

The table 4 shows the results of the calculation of the corrections
(70),(71), and (72)   for the
$(1s)^{2}2p_{\frac{1}{2}}$ state (for the $(1s)^{2}2s$ states these
corrections are equal to zero), expressed in terms of the function
$Q(\alpha Z)$ defined by
\begin{eqnarray}
\Delta E^{int}\equiv \Delta E_{c}^{(int)}+\Delta E_{tr(1)}^{(int)}+
\Delta E_{tr(2)}^{(int)}= -\frac{2^{9}}{3^{8}}\frac{m^{2}}{M}(\alpha Z)^{2}Q
(\alpha Z)\,.
\end{eqnarray}
Here we have taken into account the known non-relativistic limit
of this correction [34]. Within the $\frac{m^{2}}{M}(\alpha Z)^{4}$
approximation the function  $Q(\alpha Z)$ that we denote by
 $Q_{L}(\alpha Z)$
  is [25]
\begin{eqnarray}
Q_{L}(\alpha Z)&=&1+(\alpha Z)^{2}\Bigl(-\frac{29}{48}+\ln{\frac{9}{8}}\Bigr)
\,.
\end{eqnarray}
For comparison, this function
 is  given in the table as well.
 The functions $Q_{c}(\alpha Z)$,
 $Q_{tr(1)}(\alpha Z)$, and  $Q_{tr(2)}(\alpha Z)$
 correspond to
the corrections $\Delta E_{c}^{(int)}$, $\Delta E_{tr(1)}^{(int)}$, and
$\Delta E_{tr(2)}^{(int)}$, respectively.
In leading orders in $\alpha Z$ they are
\begin{eqnarray}
Q_{c}(\alpha Z)&=&1+(\alpha Z)^{2}\Bigl(\frac{55}{48}+\ln{\frac{9}{8}}\Bigr)
\,,\\
Q_{tr(1)}(\alpha Z)&=&-\frac{7}{4}(\alpha Z)^{2}\,,\\
Q_{tr(2)}(\alpha Z)&=&\frac{49}{64}(\alpha Z)^{4}\,.
\end{eqnarray}

For low $Z$, in addition to the corrections considered here,
the  Coulomb electron-electron interaction corrections  to the
non-relativistic nuclear recoil contribution must be calculated
separately.
The main contribution from these corrections is of order
$\frac{1}{Z}(\alpha Z)^{2}\frac{m^{2}}{M}$.

Sometimes, to estimate the nuclear recoil corrections for
high $Z$ the non-relativistic nuclear recoil operator
is averaged with the Dirac wavefunctions. But, as one can
see from the formulas (75)-(77) and the table 4, like the
one electron case (see the formulas (37)-(40)), this contribution
is considerably cancelled by the one-transverse-photon
contribution.

According to [35] the experimental value of the energy of the
$(1s)^{2}2p_{\frac{1}{2}}-(1s)^{2}2s$ transition in lithium-like
uranium is $280.59(10)\,eV$.
Let us find the total nuclear recoil contribution to the energy of this
 transition. According to our calculation  the
term $\Delta E^{int}$ is $-0.03\,eV$. Adding to this value the one-electron
contribution defined by (68) we find
$$
\Delta E_{(1s)^{2}2p_{\frac{1}{2}}}-\Delta E_{(1s)^{2}2s}=-0.07\,eV.
$$
This correction, largely made up of the  QED contributions,
 is comparable with the uncertainty of the experimental
value and, hence, will be important for
  comparison of  theory   with experiment, when
calculations of all diagrams in the second
order in $\alpha$ are completed.

\section*{Acnowledgements}
We wish to thank I.B.Khriplovich and S.G.Karshenboim for stimulating
discussions and
 K.Pachucki for making
the results of [10]  available for us prior to publication.
 The research described in this
publication was made possible in part by Grant No. NWU000  from the
International Science Foundation and Grant No. 95-02-05571a from
the Russian Foundation for Fundamental Investigations.
A.N.A. thanks I.V.Konovalov for financial support.
%%%%%%%%%%%%%%%%%%%%%%%%%%%%%%%%%%%%%%%%%%%%%%%%%%%%%%%%%%%%%%%%%%%%%%%
\newpage
\section*{Appendix}
The integration over angles in the expressions considered here
 is carried out using
the  formula
\begin{eqnarray}
\lefteqn{\sum_{m_{2}}\langle n_{1}j_{1}l_{1}
m_{1}|{\bf A}|n_{2}j_{2}l_{2}m_{2}\rangle
\langle n_{2}j_{2}l_{2}
m_{2}|{\bf B}|n_{1}j_{1}l_{1}m_{1}\rangle}\nonumber\\
& &=(-1)^{j_{1}+j_{2}-2m_{1}}\frac{1}{2j_{1}+1}
(n_{1}j_{1}l_{1}||A^{1}||n_{2}j_{2}l_{2})(n_{2}j_{2}l_{2}
||B^{1}||n_{1}j_{1}l_{1})\,,
\end{eqnarray}
where $(n_{1}j_{1}l_{1}||
A^{1}||n_{2}j_{2}l_{2})\,,\;\;(n_{2}j_{2}l_{2}||B^{1}||n_{1}j_{1}l_{1})$
are the reduced matrix elements [36].
For ${\bf A}=\mbox{\boldmath$\alpha$}\phi(r)\,,\;\;{\bf n}\phi(r)$ one
can find
\begin{eqnarray}
\lefteqn{(n_{1}j_{1}l_{1}||\mbox{\boldmath$\alpha$}\phi(r)||
n_{2}j_{2}l_{2})
=(-1)^{j_{1}-\frac{1}{2}}i\sqrt{6}\sqrt{2j_{1}+1}
\sqrt{2j_{2}+1}}\nonumber\\
& &\times \Bigl [
(-1)^{l_{1}}\delta_{l_{1}l_{2}'}
\left\{{j_{1}\,j_{2}\,1}\atop{\frac{1}{2}\,\,\frac{1}{2}\,\,l_{1}}\right\}
\int_{0}^
{\infty}g_{n_{1}j_{1}l_{1}}(r)f_{n_{2}j_{2}l_{2}}(r)
\phi(r) r^{2}dr\nonumber\\
& &-(-1)^{l_{1}'}\delta_{l_{1}'l_{2}}
\left\{{j_{1}\,j_{2}\,1}\atop{\frac{1}{2}\,\,\frac{1}{2}\,\,l_{1}'}\right\}
\int_{0}^
{\infty}f_{n_{1}j_{1}l_{1}}(r)g_{n_{2}j_{2}l_{2}}(r)
\phi(r) r^{2}dr \Bigr ]\,,
\end{eqnarray}
\begin{eqnarray}
(n_{1}j_{1}l_{1}||{\bf n}\phi(r)||
n_{2}j_{2}l_{2})
&=&(-1)^{j_{2}-\frac{1}{2}} \Bigl [
Z_{l_{1}l_{2}}^{j_{1}j_{2}}
\int_{0}^
{\infty}g_{n_{1}j_{1}l_{1}}(r)g_{n_{2}j_{2}l_{2}}(r)
\phi(r) r^{2}dr\nonumber\\
& &+Z_{l_{1}'l_{2}'}^{j_{1}j_{2}}
\int_{0}^
{\infty}f_{n_{1}j_{1}l_{1}}(r)f_{n_{2}j_{2}l_{2}}(r)
\phi(r) r^{2}dr \Bigr ]\,,
\end{eqnarray}
where
\begin{eqnarray}
Z_{l_{1}l_{2}}^{j_{1}j_{2}}=\sqrt{(2l_{1}+1)(2l_{2}+1)(2j_{1}+1)(2j_{2}+1)}
\left({l_{1}\,1\,l_{2}}\atop {0\,\,0\,0}\right)
\left\{{j_{1}\,1\,j_{2}}\atop{l_{2}\,\frac{1}{2}\,\,l_{1}}\right\}\,,
\end{eqnarray}
$l'=2j-l$;
$g_{njl}(r)$ and $f_{njl}(r)$ are the upper and lower radial components
of the Dirac wavefunction [37]:
$$
\psi_{njlm}({\bf r})=
\left(\begin{array}{c}
g_{njl}(r)\Omega_{jlm}({\bf n})\\
if_{njl}(r)\Omega_{jl'm}({\bf n})
\end{array}\right)\;.\\
$$

\newpage
\begin{table}
\caption{The  results of the numerical calculation of the
one-electron nuclear
recoil corrections to the $1s$ state energy
expressed in terms of the function $P(\alpha Z)$ defined by equation (56).
 $P_{S}(\alpha Z)$ is the
Salpeter's contribution defined by equation (57).}
\begin{tabular}{|c|l|l|l|l|l|}  \hline
$Z$&$P_{c}(\alpha Z)$&$P_{tr(1)}(\alpha Z)$&$P_{tr(2)}(\alpha Z)$&
$P(\alpha Z)$&$P_{S}(\alpha Z)$\\ \hline
1&-1.3111(2)&12.568(2)&-5.8267(3)&5.430(2)&5.4461\\ \hline
5&-1.2345(1)&8.5854(3)&-3.0476(2)&4.3033(4)&4.3731\\ \hline
10&-1.1586&6.9974(1)&-2.0438&3.7950(1)&3.9110\\ \hline
15&-1.0994&6.1340(1)&-1.5373&3.4973(1)&3.6407\\ \hline
20&-1.0537&5.5678(1)&-1.2201&3.2940(1)&3.4489\\ \hline
25&-1.0192&5.1671(1)&-0.9996&3.1483(1)&3.3001\\ \hline
30&-0.9946&4.8744(1)&-0.8362&3.0437(1)&3.1786\\ \hline
35&-0.9790&4.6598(1)&-0.7094&2.9714(1)&3.0758\\ \hline
40&-0.9721&4.5065(1)&-0.6076&2.9268(1)&2.9868\\ \hline
45&-0.9740&4.4048(1)&-0.5231&2.9077(1)&2.9083\\ \hline
50&-0.9849&4.3496(1)&-0.4510&2.9137(1)&2.8380\\ \hline
55&-1.0059&4.3389(1)&-0.3874&2.9456(1)&2.7745\\ \hline
60&-1.0383&4.3739(2)&-0.3295&3.0061(2)&2.7165\\ \hline
65&-1.0845(1)&4.4588(2)&-0.2746&3.0997(2)&2.6631\\ \hline
70&-1.1479(2)&4.6014(3)&-0.2201&3.2334(4)&2.6137\\ \hline
75&-1.2339(3)&4.8153(7)&-0.1631&3.4183(8)&2.5677\\ \hline
80&-1.3506(5)&5.122(1)&-0.0996(1)&3.672(1)&2.5247\\ \hline
85&-1.512(1)&5.558(4)&-0.0237(2)&4.022(4)&2.4843\\ \hline
90&-1.741(3)&6.186(7)&0.0743(9)&4.519(8)&2.4462\\ \hline
92&-1.861(5)&6.51(1)&0.123(1)&4.77(1)&2.4315\\ \hline
95&-2.084(9)&7.12(3)&0.212(1)&5.25(3)&2.4101\\ \hline
100&-2.64(3)&8.6(1)&0.428(6)&6.4(1)&2.3759\\ \hline
\end{tabular}
\end{table}

\newpage
\begin{table}
\caption{The  results of the numerical calculation of the
one-electron nuclear
recoil corrections to the $2s$ state energy
expressed in terms of the function $P(\alpha Z)$ defined by equation (56).
 $P_{S}(\alpha Z)$ is the
Salpeter's contribution defined by equation (58).}
\begin{tabular}{|c|l|l|l|l|l|}  \hline
$Z$&$P_{c}(\alpha Z)$&$P_{tr(1)}(\alpha Z)$&$P_{tr(2)}(\alpha Z)$&
$P(\alpha Z)$&$P_{S}(\alpha Z)$\\ \hline
1&-1.3112(2)&13.177(1)&-5.7103(3)&6.155(1)&6.1710\\ \hline
5&-1.2351(1)&9.1911(2)&-2.9225(1)&5.0335(2)&5.0980\\ \hline
10&-1.1612&7.6075(1)&-1.9080&4.5383(1)&4.6359\\ \hline
15&-1.1055&6.7562&-1.3908&4.2599&4.3656\\ \hline
20&-1.0647&6.2093&-1.0621&4.0825&4.1738\\ \hline
25&-1.0367&5.8352&-0.8294&3.9691&4.0251\\ \hline
30&-1.0202&5.5767&-0.6528&3.9037&3.9035\\ \hline
35&-1.0147&5.4047&-0.5115&3.8785&3.8008\\ \hline
40&-1.0202&5.3037&-0.3935&3.8900&3.7117\\ \hline
45&-1.0372&5.2656&-0.2908&3.9376&3.6332\\ \hline
50&-1.0668&5.2876(1)&-0.1980&4.0228(1)&3.5630\\ \hline
55&-1.1108&5.3711(1)&-0.1105&4.1498(1)&3.4994\\ \hline
60&-1.1723(1)&5.5218(1)&-0.0247&4.3248(2)&3.4414\\ \hline
65&-1.2554(1)&5.7504(2)&0.0634&4.5584(2)&3.3881\\ \hline
70&-1.3668(2)&6.0743(4)&0.1581(1)&4.8656(5)&3.3387\\ \hline
75&-1.5164(4)&6.5211(7)&0.2651(1)&5.2698(8)&3.2927\\ \hline
80&-1.7199(7)&7.135(2)&0.3921(2)&5.807(2)&3.2496\\ \hline
85&-2.003(1)&7.988(4)&0.5516(4)&6.537(4)&3.2092\\ \hline
90&-2.413(4)&9.205(8)&0.7645(6)&7.557(9)&3.1711\\ \hline
92&-2.630(7)&9.84(1)&0.872(1)&8.08(2)&3.1565\\ \hline
95&-3.04(2)&11.02(2)&1.070(2)&9.05(3)&3.1351\\ \hline
100&-4.07(5)&13.9(1)&1.55(1)&11.4(2)&3.1009\\ \hline
\end{tabular}
\end{table}
\newpage
\begin{table}
\caption{The  results of the numerical calculation of the
one-electron nuclear
recoil corrections to the $2p_{\frac{1}{2}}$ state energy
expressed in terms of the function $P(\alpha Z)$ defined by equation (56).
 $P_{S}(\alpha Z)$ is the
Salpeter's contribution defined by equation (59).}
\begin{tabular}{|c|l|l|l|l|l|}  \hline
$Z$&$P_{c}(\alpha Z)$&$P_{tr(1)}(\alpha Z)$&$P_{tr(2)}(\alpha Z)$&
$P(\alpha Z)$&$P_{S}(\alpha Z)$\\ \hline
   1&-0.0000&-0.1440&-0.1571&-0.3011&-0.3088\\ \hline
   5&-0.0007&-0.1492&-0.1194&-0.2692&-0.3088\\ \hline
  10&-0.0024&-0.1526&-0.0727&-0.2277&-0.3088\\ \hline
  15&-0.0051&-0.1535&-0.0258&-0.1845&-0.3088\\ \hline
  20&-0.0088&-0.1524&0.0218&-0.1393&-0.3088\\ \hline
  25&-0.0133&-0.1493&0.0706&-0.0920&-0.3088\\ \hline
  30&-0.0189&-0.1444&0.1212&-0.0421&-0.3088\\ \hline
  35&-0.0255&-0.1375&0.1742&0.0112&-0.3088\\ \hline
  40&-0.0335&-0.1284&0.2304&0.0685&-0.3088\\ \hline
  45&-0.0432&-0.1165&0.2906&0.1310&-0.3088\\ \hline
  50&-0.0548&-0.1012&0.3560&0.2000&-0.3088\\ \hline
  55&-0.0691&-0.0814&0.4278&0.2774&-0.3088\\ \hline
  60&-0.0868&-0.0555&0.5078&0.3655&-0.3088\\ \hline
  65&-0.1091&-0.0211&0.5982&0.4680&-0.3088\\ \hline
  70&-0.1376&0.0252&0.7018&0.5894&-0.3088\\ \hline
  75&-0.1750&0.0891(1)&0.8229&0.7370(1)&-0.3088\\ \hline
  80&-0.2253(1)&0.1796(1)&0.9671&0.9214(2)&-0.3088\\ \hline
  85&-0.2954(2)&0.3123(3)&1.1429(1)&1.1598(4)&-0.3088\\ \hline
  90&-0.3972(6)&0.515(1)&1.3632(1)&1.481(1)&-0.3088\\ \hline
  92&-0.451(1)&0.626(1)&1.468(2)&1.643(3)&-0.3088\\ \hline
  95&-0.554(2)&0.842(3)&1.649(3)&1.937(5)&-0.3088\\ \hline
 100&-0.816(9)&1.41(1)&2.040(3)&2.63(2)&-0.3088\\ \hline
\end{tabular}
\end{table}
\newpage
\begin{table}
\caption{The  results of the numerical calculation of the
two-electron nuclear
recoil corrections $\Delta E^{(int)}$ for the $(1s)^{2}2p_{\frac{1}{2}}$
 state  of lithium-like ions
expressed in terms of the function $Q(\alpha Z)$ defined by equation (74).
 $Q_{L}(\alpha Z)$ is the leading contribution
defined by equation (75).}
\begin{tabular}{|c|l|l|l|l|l|}  \hline
$Z$&$Q_{c}(\alpha Z)$&$Q_{tr(1)}(\alpha Z)$&$Q_{tr(2)}(\alpha Z)$&
$Q(\alpha Z)$&$Q_{L}(\alpha Z)$\\ \hline
   5&1.00168&-0.00233&0.00000&0.99935&0.99935\\ \hline
  10&1.00677&-0.00938&0.00002&0.99741&0.99741\\ \hline
  15&1.01533&-0.02129&0.00011&0.99416&0.99417\\ \hline
  20&1.02753&-0.03830&0.00036&0.98959&0.98964\\ \hline
  25&1.04359&-0.06077&0.00088&0.98370&0.98381\\ \hline
  30&1.06378&-0.08920&0.00186&0.97645&0.97669\\ \hline
  35&1.08851&-0.12422&0.00353&0.96782&0.96827\\ \hline
  40&1.11827&-0.16669&0.00617&0.95776&0.95856\\ \hline
  45&1.15370&-0.21767&0.01019&0.94622&0.94755\\ \hline
  50&1.19560&-0.27853&0.01607&0.93313&0.93525\\ \hline
  55&1.24500&-0.35105&0.02447&0.91841&0.92165\\ \hline
  60&1.30322&-0.43751&0.03625&0.90195&0.90676\\ \hline
  65&1.37198&-0.54091&0.05254&0.88361&0.89057\\ \hline
  70&1.45352&-0.66521&0.07488&0.86320&0.87309\\ \hline
  75&1.55087&-0.81573&0.10538&0.84052&0.85431\\ \hline
  80&1.66810&-0.99980&0.14699&0.81529&0.83424\\ \hline
  85&1.81092&-1.22771&0.20395&0.78716&0.81287\\ \hline
  90&1.98751&-1.51431&0.28250&0.75570&0.79021\\ \hline
  92&2.07014&-1.65003&0.32196&0.74206&0.78078\\ \hline
  95&2.21001&-1.88186&0.39221&0.72035&0.76625\\ \hline
 100&2.49719&-2.36503&0.54826&0.68041&0.74099\\ \hline
\end{tabular}
\end{table}

\newpage
\begin{thebibliography}{33}
\bibitem{s1}
H.A.Bethe and E.E.Salpeter, Quantum Mechanics of One- and Two-Electron
Atoms (Springer, Berlin, 1957).
\bibitem{s2}
E.E.Salpeter and H.A.Bethe, Phys.Rev. {\bf 84}, 1232 (1951).
\bibitem{s3}
E.E.Salpeter, Phys.Rev. {\bf 87}, 328 (1952).
\bibitem{s4}
T.Fulton and P.C.Martin, Phys.Rev. {\bf 95}, 811 (1954).
\bibitem{s5}
H.Grotch and D.R.Yennie, Rev.Mod.Phys. {\bf 41}, 350 (1969).
\bibitem{s6}
G.W.Erickson and D.R.Yennie, Ann.Phys. (NY) {\bf 35}, 271 (1965);
G.W.Erickson, in:  Physics of One- and Two-Electron Atoms, eds.
F.Bopp and H.Kleinpoppen (North-Holland, Amsterdam, 1970).
\bibitem{s7}
M.Doncheski, H.Grotch and G.W.Erickson, Phys.Rev.A {\bf 43}, 2152 (1991).
\bibitem{s8}
I.B.Khriplovich, A.I.Milstein and A.S.Yelkhovsky,
 Phys.Scr. T {\bf 46}, 252 (1993).
\bibitem{s9}
R.N.Fell, I.B.Khriplovich, A.I.Milstein
 and A.S.Yelkhovsky, Phys.Lett.A {\bf 181},
172 (1993).
\bibitem{s10}
K.Pachucki and H.Grotch, Phys.Rev.A (to be published).
\bibitem{s11}
E.A.Golosov, I.B.Khriplovich, A.I.Milstein, and A.S.Yelkhovsky,
Zh.Eksp.Teor.Fiz. {\bf 107}, 393 (1995).
\bibitem{s12}
L.N.Labzowsky, In: Papers at 17th All-Union Symposium on Spectroscopy
(Astrosovet, Moscow, 1972), Part 2, pp. 89-93.
\bibitem{s13}
M.A.Braun, Zh.Eksp.Teor.Fiz. {\bf 64}, 413 (1973).
\bibitem{s14}
V.M.Shabaev, Teor.Mat.Fiz. {\bf 63}, 394 (1985)
(Theor.Math.Phys. {\bf 63}, 588
(1985)).
\bibitem{s15}
V.M.Shabaev, In: Papers at First Soviet-British Symposium on Spectroscopy
of Multicharged Ions (Academy of Sciences, Troitsk, 1986), pp. 238-240.
\bibitem{s16}
V.M.Shabaev, Yad.Fiz. {\bf 47}, 107 (1988) (Sov.J.Nucl.Phys. {\bf 47} 69
(1988)).
\bibitem{s17}
L.S.Dul'yan and R.N.Faustov, Teor.Mat.Fiz. {\bf 22}, 314 (1975).
\bibitem{s18}
F.Gross, Phys.Rev. {\bf 186}, 1448 (1969).
\bibitem{19}
A.A.Logunov and A.N.Tavkhelidze, Nuovo Cimento {\bf 29}, 380 (1963).
\bibitem{s20}
R.N.Faustov, Fiz.Elem.Chast.At.Yad. {\bf 3}, 238 (1972).
\bibitem{s21}
G.P.Lepage, Phys.Rev.A {\bf 16},863 (1977).
\bibitem{s22}
V.M.Shabaev, In: Many-Particle Effects in Atoms,
 ed. U.I.Safronova (Academy of Sciences,
Moscow, 1985), pp. 118-144.
\bibitem{s23}
G.Breit, Phys.Rev. {\bf 35}, 1447 (1930).
\bibitem{s24}
A.S.Yelkhovsky, Preprint BINP 94-27
(Budker Inst. of Nuclear Physics, Novosibirsk,
1994).
\bibitem{s25}
V.M.Shabaev and A.N.Artemyev, J.Phys.B {\bf 27}, 1307 (1994).
\bibitem{s26}
C.W.Palmer, J.Phys.B {\bf 20}, 5987 (1987)
\bibitem{s27}
J.Epstein and S.Epstein, Am.J.Phys. {\bf 30}, 266 (1962).
\bibitem{s28}
V.M.Shabaev, Vestn.Leningrad.Univ. N4, 15 (1984).
\bibitem{s29}
V.M.Shabaev, J.Phys.B {\bf 24}, 4479 (1991).
\bibitem{s30}
I.P.Grant and H.M.Quiney, Adv.At.Mol.Phys. {\bf 23}, 37 (1988).
\bibitem{s31}
W.R.Johnson, S.A.Blundell, and J.Sapirstein, Phys.Rev.A {\bf 37}, 307 (1988).
\bibitem{s32}
S.Salomonson and P.\"Oster, Phys.Rev.A {\bf 40}, 5548 (1989).
\bibitem{s33}
C.Froese Fisher and F.A.Parpia, Phys.Lett.A {\bf 179}, 198 (1993).
\bibitem{s34}
D.S.Hughes and C.Eckart, Phys.Rev. {\bf 36}, 694 (1930).
\bibitem{s35}
J.Schweppe, A.Belkacem, L.Blumenfeld, N.Claytor, B.Feinberg,
H.Gould, V.E.Kostroun, L.Levy, S.Misawa, J.R.Mowat, and
M.H.Prior, Phys.Rev.Lett. {\bf 66}, 1434 (1991).
\bibitem{s36}
I.I.Sobel'man, Introduction to  Theory of Atomic Spectra
(Nauka, Moscow, 1977).
\bibitem{s37}
A.I.Akhiezer and V.B.Berestetsky, Quantum Electrodynamics
 (Nauka, Moscow, 1969).
\end{thebibliography}
\end{document}